\documentclass[11pt,a4paper]{article}

\usepackage{amssymb,latexsym, amsfonts}
\usepackage{mathrsfs,amsmath,amsthm,bm}
\usepackage{graphicx}

\DeclareMathOperator{\Tr}{Tr}
\newcommand{\pd}{\partial}
\newcommand{\defeq}{\stackrel{\rm def}{=}}
\newcommand{\mc}{\mathcal}
\newcommand{\mb}{\mathbb}
\newcommand{\bs}{\boldsymbol}
\def\defeq{\stackrel{\rm def}{=}}

\def\gij,k{\Gamma_{ij,k}}
\def\gki,j{\Gamma_{ki,j}}
\def\gkj,i{\Gamma_{kj,i}}
\def\gij,k{\Gamma_{ij,k}}

\def\gik,j{\Gamma_{ik,j}}

\def\argmin{\mathop{\rm arg\,min}}

\def\be{\begin{equation}}
\def\ee{\end{equation}}
\def\bea{\begin{eqnarray}}
\def\eea{\end{eqnarray}}
\def\bt{\begin{theorem}}
\def\et{\end{theorem}}
\def\bd{\begin{definition}}
\def\ed{\end{definition}}
\def\bfig{\begin{figure}}
\def\efig{\end{figure}}
\def\bc{\begin{center}}
\def\ec{\end{center}}

\def\nn{\nonumber}

\begin{document}

\title{Mean-field equations for higher-order quantum statistical models : an information geometric approach}

\author{
  N Yapage\footnote{\tt nihal@maths.ruh.ac.lk}\\
  Department of Mathematics\\ University of Ruhuna, \\
  Matara\\ Sri Lanka.
}
\maketitle

\begin{abstract}
 This work is a simple extension of \cite{NNjpa}. We apply the concepts of information geometry to study the mean-field approximation for a general class of quantum statistical models namely the higher-order quantum Boltzmann machines (QBMs).  The states we consider are assumed to have at most third-order interactions with deterministic coupling coefficients. Such states, taken together, can be shown to form a quantum exponential family and thus can be viewed as a smooth manifold. In our work, we explicitly obtain naive mean-field equations for the third-order classical and quantum Boltzmann machines and demonstrate how some information geometrical concepts, particularly, exponential and mixture projections used to study the naive mean-field approximation in \cite{NNjpa} can be extended to a more general case. Though our results do not differ much from those in \cite{NNjpa}, we emphasize the validity and the importance of information geometrical point of view for higher dimensional classical and quantum statistical models.
\end{abstract}

{\bf Keywords:} mean-field theory, quantum statistical model, information geometry, quantum relative entropy, quantum exponential family

\section{Introduction}
The mean-field approximation uses a simple tractable family of density operators to calculate quantities related to a complex density operator including mutual interactions. Information geometry, on the other hand, studies intrinsic geometrical structure existing in the manifold of density operators     \cite{ANamsox}. Many authors have used mean-field approximation to classical statistical models like classical Boltzmann machines (CBMs) \cite{MDmit} and also have discussed the properties in the in the information geometrical point of view \cite{Tnc,AKNnn}. In this work, we apply mean-field theory to the third-order CBMs and QBMs and derive the naive mean-field equations using information geometrical concepts.

\section{Information geometry of mean-field approximation for third-order CBMs}
Let us  consider a network of $n$ elements numbered as $1,2,\ldots,n$. Let the value 
of each element $i\in\{1,2,\ldots,n\}$ be $x_i\in \{-1,+1\}$. Then 
a state of the network can be represented as 
$\bs{x}=(x_1,x_2,\ldots,x_n)\in\{-1,+1\}^n$. Each element  
$i\in\{1,2,\ldots,n\}$ carries a threshold value $\theta_i \in {\mb R}$. The network also has 
a real-valued parameter $w_{ij}$ for each pair of elements $\{i,j\}$, 
which is called the coupling coefficient between $i$ and $j$. 
These parameters are assumed to satisfy conditions 
$w_{ij}=w_{ji},\; w_{ii}=0$. The other real-valued parameter is $v_{ijk}$ and is symmetric on all pairs of indices.
The equilibrium (stationary) distribution is given by the probability distributions of the form 
\bea
p({\bs x},h,w,v)
&=&\exp\Bigl\{\sum_i h_i x_i 
+ \sum_{i<j} w_{ij} x_i x_j   
+\sum_{i<j<k} v_{ijk} x_i x_j x_k-\psi(h,w,v)\Bigr\} 
\label{eq:3rdorderCBM} 
\eea 
with
\be 
\psi(h,w,v) = 
\log \sum_{\bs x}\exp\Bigl\{\sum_i h_i x_i 
+ \sum_{i<j} w_{ij} x_i x_j   
+\sum_{i<j<k} v_{ijk} x_i x_j x_k\Bigr\},
\ee
where ${\bs x}=(x_1, \ldots, x_n)\in\{-1,+1\}^n$.
Thus, noting that the correspondence $p_{h,w,v}\leftrightarrow (h,w,v)$ 
is one to one, we can, at least mathematically, identify each third-order CBM \cite{TJS} with its equilibrium probability distribution.

Many good properties of such networks are consequences of the fact 
that the equilibrium distributions form an exponential family. 
Here, we discuss this important aspect of the CBM \cite{NKcs} briefly.
Let ${\mc X}$ be a finite set or, more generally, a measurable space with 
an underlying measure ${\rm d}\mu$. We denote the set of positive probability
distributions (probability mass functions for a finite ${\mc X}$ and 
probability density functions for a general $({\mc X}, {\rm d}\mu)$) 
on ${\mc X}$ by ${\mc P}=\mc{P(X)}$. When a family of distributions, say 
\be
{\mc M} = \{p_{\theta}\:\vert\: \theta = (\theta^i);\,\, i=1,\ldots,n \}
\subset {\mc P},
\ee 
is represented in the form
\be
p_{\theta}(x) = \exp \Bigl\{c(x) + \sum_{i=1}^n \theta^i f_i (x)
- \psi(\theta)\Bigr\},\quad x\in{\mc X}, \label{eq:cexpfam}
\ee
${\mc M}$ is called an exponential family. 
Here, $\theta^i;\,i=1,\ldots,n$ are $\mb R$-valued parameters, $c$ and $f_i$ are 
functions on ${\mc X}$ and $\psi(\theta)$ is a real-valued convex function.   
 Further, we assume that the 
correspondence $\theta \mapsto p_\theta$  is one to one. These 
$\theta = (\theta^i)$ are called the natural coordinates of $\mc M$. 
 
Now, for the exponential family $\mc M$, if we let
\[
\eta_i(\theta) \defeq  {\mb E}_{\theta}[f_i] 
        = \sum_{\bs x} p_\theta (\bs x ) f_i (\bs x )
\]     
then  $\eta = (\eta_i)$ and $\theta = (\theta^i)$ are in one-to-one 
correspondence. That is, 
we can also use $\eta$ instead of $\theta$ to specify an element of 
$\mc M$. These $(\eta_i)$ are called the expectation coordinates
of $\mc M$. The expectation coordinates are, in general, represented 
as 
\be
\eta_i = \pd_i \psi (\theta )\qquad \Bigl(\pd_i \defeq 
\frac{\pd}{\pd\theta^i} \Bigr).\label{eq:expect-param}
\ee

The set that consists of equilibrium probability distributions of third-order CBM  (\ref{eq:3rdorderCBM}) is one example of exponential family. 
In addition, threshold values, coupling 
coefficients (weights) and third-oder weights become the natural coordinates while 
${\mb E}_\theta [x_i ]$, ${\mb E}_\theta [x_i x_j ]$ and 
${\mb E}_\theta [x_i x_j x_k ]$ become expectation 
coordinates. The notion of exponential family is very important in 
statistics and information geometry, and is also useful in studying properties 
of third-order CBMs with their mean-field approximations.

We now consider a hierarchy of exponential families. Let ${\mc P}_r$ be the set of $r$th-order CBMs. Then ${\mc P}_r$ also turns out to be an exponential family.  Thus, we have a hierarchical structure of exponential families
${\mc P}_1 \subset {\mc P}_2 \subset \cdots \subset {\mc P}_r$.  
In particular, ${\mc P}_1,{\mc P}_2$ and ${\mc P}_3$ can be represented by
\bea
{\mc P}_1 
&=& \{p_1 (x_1) \cdots p_n (x_n)\}=\{\mbox{product distributions}\}\nn, \\
{\mc P}_2 
&=& \{\mbox{equilibrium distributions of CBM}\}\\
\quad\mbox{and}\nn \\
{\mc P}_3 
&=& \{\mbox{equilibrium distributions of third-order CBM}\}
\eea 
respectively. 

In this subsection, we derive the naive mean-field equation for third-order CBM. 
When the system size is large, the partition function $\exp(\psi(h,w,v))$ 
is very difficult to calculate and thus explicit calculation of the 
expectations $m_i$ is intractable. Therefore, due to that difficulty, we are led to obtain a good approximation of $m_i$ for a given probability distribution $p_{h,w,v}\in {\mc P}_3$.

First, we consider the subspace ${\mc P}_1$ of ${\mc P}_3$. 
We parametrize each distribution in ${\mc P}_1$ by $\bar{h}$
and write as 
\be
p_{\bar h}(\bm x) = \exp\Big\{\sum_i {\bar h}_i x_i -\psi(\bar h)\Big\},
\ee
where 
\[
\psi(\bar h) = \sum_i \log\Bigl\{ \exp({\bar{h}}_i) + 
\exp(-{\bar{h}}_i)\Bigr\}.
\] 
Then,  ${\mc P}_1$ forms a submanifold of ${\mc P}_3$ specified by $w_{ij}=0=v_{ijk}$ and ${\bar{h}}_i$ as its coordinates. 
The expectations ${\bar{m}}_i := {\mb E}_{\bar{h}} [x_i]$ 
form another coordinate system of ${\mc P}_1$. For a given 
$p_{\bar{h}} \in {\mc P}_1$, it is easy to obtain 
${\bar{m}}_i = {\mb E}_{\bar{h}} [x_i]$  from ${\bar{h}}_i$ because 
$x_{i}$'s are independent. We can calculate ${\bar{m}}_i$ to be
\be
{\bar{m}}_i =  \frac{\pd \psi(\bar{h})}{\pd {\bar{h}}_i}  
=\frac{\exp({\bar{h}}_i)-\exp(-{\bar{h}}_i)}
{\exp({\bar{h}}_i)+\exp(-{\bar{h}}_i)}
=\tanh({\bar{h}}_i),\label{eq:meanpara}
\ee
from which we obtain
\be
 {\bar{h}}_i = \frac{1}{2}
\log\biggl(\frac{1+{\bar{m}}_i}{1-{\bar{m}}_i}\biggr).\label{eq:hibar}
\ee
The simple idea behind the mean field approximation for a 
$p_{h,w,v}\in {\mc P}_3$ is to use quantities obtained in the 
form of expectation with respect to some relevant 
$p_{\bar{h}} \in {\mc P}_1$. 

Now, we need a suitable criterion to measure the approximation 
of two probability distributions $q\in {\mc P}$ and $p_{\theta} \in {\mc M}$. For the present purpose, we adopt the Kullback-Leibler (KL) divergence (relative entropy) 
\be
  D(q\| p_{\theta}) \defeq \sum_{\bs x}
        q({\bs x})\log \frac{q({\bs x})}{p_{\theta}({\bs x})}\label{eq:kldiv}
\ee.  
Given $p_{h,w,v}\in {\mc P}_3$, its e-(exponential) and m-(mixture) 
projections (see \cite{ANamsox}) onto ${\mc P}_1$ are defined by 
\be
{\bar{p}}^{(e)} = p_{\bar{h}^{(e)}}\defeq 
\argmin_{p_{\bar{h}} \in {\mc P}_1} D(p_{\bar{h}}\|p_{\theta})\label{eq:eproj1}
\ee
and
\be
{\bar{p}}^{(m)} = p_{\bar{h}^{(m)}}\defeq 
\argmin_{p_{\bar{h}} \in {\mc P}_1} D(p_{\theta}\|p_{\bar{h}})\label{eq:mproj1}
\ee
respectively, where 
\be
\bar{h}^{(e)} = \argmin_{\bar{h}=({\bar{h}}_i)} D(p_{\bar{h}}\|p_{\theta})
\label{eq:eproj}
\ee
and
\be
\bar{h}^{(m)} =  \argmin_{\bar{h}=({\bar{h}}_i)} D(p_{\theta}\|p_{\bar{h}}).
\label{eq:mproj}
\ee
As necessary conditions, we have 
\be
\frac{\pd}{\pd {\bar{h}}_i} D(p_{\bar{h}}\|p_{\theta})= 0 \label{eq:eproject}
\ee
and
\be
\frac{\pd}{\pd {\bar{h}}_i} D(p_{\theta}\|p_{\bar{h}}) = 0,\label{eq:mproject}
\ee 
which are weaker than (\ref{eq:eproj}) and (\ref{eq:mproj}). But sometimes 
(\ref{eq:eproject}) and (\ref{eq:mproject}) are chosen to be the definitions 
of e-, m- projections respectively for convenience. 
It can be shown that the m-projection ${\bar{p}}^{(m)}$ 
gives the true values of expectations, that is 
$m_i = {\bar{m}}_i$ or ${\mb E}_{(\theta)} [x_i] = {\mb E}_{\bar{h}} [x_i]$
for $p_{\bar{h}} = {\bar{p}}^{(m)}$. The e-projection ${\bar{p}}^{(e)}$ from ${\mc P}_3$ onto ${\mc P}_1$ gives 
the naive mean-field approximation for third-order CBM. 
Now we derive the naive mean-field equation for third-order CBM following \cite{Tnc}. 
Recall that the equilibrium distribution for third-order CBM is given by
\bea
p\defeq p({\bs x},h,w,v)
&=&\exp\Bigl\{\sum_i h_i x_i 
+ \sum_{i<j} w_{ij} x_i x_j   
+\sum_{i<j<k} v_{ijk} x_i x_j x_k-\psi(h,w,v)\Bigr\} 
\label{eq:3rdordercbm} 
\eea 
with
\be 
\psi(h,w,v) = 
\log \sum_{\bs x}\exp\Bigl\{\sum_i h_i x_i 
+ \sum_{i<j} w_{ij} x_i x_j   
+\sum_{i<j<k} v_{ijk} x_i x_j x_k\Bigr\},
\ee
where ${\bs x}=(x_1, \ldots, x_n)\in\{-1,+1\}^n$.
Now we define another function 
\be
\phi(p)=\phi(h,w,v) \defeq  
\sum_{i<j<k} v_{ijk}\;{\mb E}_p[x_i x_j x_k] 
+ \sum_{i<j} w_{ij}\;{\mb E}_p[x_i x_j]
+ \sum_i h_i\;{\mb E}_p[x_i] - \psi(p),
\ee
which coincides with the negative entropy
\be 
\phi(p) = \sum_{\bs x} p(\bs x) \log p(\bs x).
\ee 
In particular, for a product distribution $p_{\bar{h}}\in{\mc P}_1$, 
using ${\mb E}_{\bar{h}} [x_i] = {\bar{m}}_i$ , we have 
\be
\phi(p_{\bar{h}}) = \sum_i\biggl[\biggl(\frac{1+{\bar{m}}_i}{2}\biggr)
\log\biggl(\frac{1+{\bar{m}}_i}{2}\biggr) 
+ \biggl(\frac{1-{\bar{m}}_i}{2}\biggr)
\log\biggl(\frac{1-{\bar{m}}_i}{2}\biggr) \biggr].\label{eq:prodent}
\ee
The KL divergence between $p\in {\mc P}_3$ and $p_{\bar{h}}\in {\mc P}_1$ 
can be expressed in the following form:
\bea
D(p_{\bar{h}}\|p) &=& \psi(p)+\phi(p_{\bar{h}})- 
\sum_{i<j<k} v_{ijk}\;{\mb E}_p[x_i x_j x_k] 
- \sum_{i<j} w_{ij}\;{\mb E}_p[x_i x_j]
- \sum_i h_i \;{\mb E}_{\bar{h}}[x_i]\nn \\
&=& \psi(p)+\phi(p_{\bar{h}})
- \sum_{i<j<k} v_{ijk} {\bar{m}}_i {\bar{m}}_j {\bar{m}}_k 
-\sum_{i<j} w_{ij} {\bar{m}}_i {\bar{m}}_j 
- \sum_i h_i {\bar{m}}_i. \nn\\
&=& \psi(p) + \frac{1}{2}\sum_i\biggl[(1+{\bar{m}}_i)
\log\biggl(\frac{1+{\bar{m}}_i}{2}\biggr) + (1-{\bar{m}}_i)
\log\biggl(\frac{1-{\bar{m}}_i}{2}\biggr) \biggr]\nn \\
&&  - \sum_{i<j<k} v_{ijk} {\bar{m}}_i {\bar{m}}_j {\bar{m}}_k 
-\sum_{i<j} w_{ij} {\bar{m}}_i {\bar{m}}_j - \sum_i h_i {\bar{m}}_i.
\label{eq:KLdiv}
\eea
Now consider the e-projection (\ref{eq:eproject}) from $p\in {\mc P}_3$ onto 
$p_{\bar{h}}\in {\mc P}_1$, i.e. 
\be
\frac{\pd}{\pd {\bar{h}}_i}D(p_{\bar{h}}\|p)= 0.
\ee
Noting that $\bar{h}$ and  $\bar{m}$ are in one-to-one correspondence, 
we may consider instead 
\be
\frac{\pd}{\pd {\bar{m}}_i} D(p_{\bar{h}}\|p) = 0.
\ee
Since $\psi(p)$ does not depend on  ${\bar{m}}_i$, we obtain from 
(\ref{eq:KLdiv}) that
\bea
0 &=& \frac{1}{2}\log\biggl(\frac{1+{\bar{m}}_i}{1-{\bar{m}}_i}\biggr) 
- \sum_{k\neq j\neq i} v_{ijk} {\bar{m}}_j {\bar{m}}_k- \sum_{j\neq i} w_{ij} {\bar{m}}_j - h_i \nn\\
 &=& {\bar{h}}_i - \sum_{k\neq j\neq i} v_{ijk} {\bar{m}}_j {\bar{m}}_k- \sum_{j\neq i} w_{ij} {\bar{m}}_j - h_i,\label{eq:stat}
\eea
where the second equality is from (\ref{eq:hibar}). 
Thus the naive mean-field equation is obtained from (\ref{eq:meanpara}) and 
(\ref{eq:stat}) as 
\be
\tanh^{-1}({\bar{m}}_i) = \sum_{k\neq j\neq i} v_{ijk} {\bar{m}}_j {\bar{m}}_k+ \sum_{j\neq i} w_{ij} {\bar{m}}_j + h_i
\ee
and this is usually written in the form 
\be
{\bar{m}}_i =  \tanh\biggl(\sum_{k\neq j\neq i} v_{ijk} {\bar{m}}_j {\bar{m}}_k +\sum_{j\neq i} w_{ij} {\bar{m}}_j + h_i\biggr).
\ee
\section{Definition of third-order quantum Boltzmann machines}
Let us consider an $n$-element system of quantum spin-half particles. 
Each element is 
represented as a quantum spin with local Hilbert space $\mb{C}^2$, and the $n$-element system corresponds to 
${\mc H} \equiv (\mb{C}^2)^{\otimes n}\simeq {\mb C}^{2^n}$. Let ${\mc S}$ be 
the set of strictly positive states on ${\mc H}$; 
\be
{\mc S} = \{\rho \:\vert\: \rho=\rho^\ast > 0 \,\,\mbox{and}\,\,\Tr\rho =1\}.
\ee
Here, each $\rho$ is a $2^n \times 2^n$ matrix; $ \rho=\rho^\ast >0$  
means that $\rho$ is Hermitian and positive definite 
respectively; and  $\Tr\rho =1$ shows that the trace of the density matrix 
$\rho$ is unity. 
Now an element of ${\mc S}$ is said to have at most $r$th-order interactions if it is written as 
\bea
\rho_{\theta}&=&\exp\Bigl\{\sum_{i,s}
       {\theta}_{is}^{(1)} \sigma_{is} + 
\sum_{i < j}\sum_{s,t}{\theta}_{ijst}^{(2)} \sigma_{is} \sigma_{jt}+ \cdots +\sum_{i_1 <\cdots< i_r}\sum_{s_1 \ldots s_r}
{\theta}_{i_1 \ldots i_r s_1 \ldots s_r}^{(r)} \sigma_{i_1 s_1} \cdots 
\sigma_{i_r s_r}-\psi(\theta)\Bigr\}\nonumber 
\eea
\be
= \exp\Bigl\{\sum_{j=1}^r 
\sum_{i_1 < \cdots <i_j} \sum_{s_1 \ldots s_j}
       {\theta}_{i_1 \ldots i_j s_1 \ldots s_j}^{(j)}
\sigma_{i_1 s_1} \cdots \sigma_{i_j s_j}-\psi(\theta)\Bigr\}
\label{eq:kelement} 
\ee
with 
\be 
\psi(\theta) = \log \Tr\exp\Bigl\{\sum_{j=1}^r 
\sum_{i_1 < \cdots <i_j} \sum_{s_1 \ldots s_j}
       {\theta}_{i_1 \ldots i_j s_1 \ldots s_j}^{(j)}
\sigma_{i_1 s_1} \cdots \sigma_{i_j s_j}\Bigr\}, 
\ee
where $\sigma_{i s}=I^{\otimes (i-1)}\otimes \sigma_{s}\otimes I^{\otimes (n-i)}$, 
$\theta = ({\theta}_{i_1 \ldots i_j s_1 \ldots s_j}^{(j)})$. Here, $I$ is the 
identity matrix on $\mc H$ and $\sigma_s$ for $s\in\{1,2,3\}$ are the usual Pauli 
matrices given by
$$
\sigma_1 = \left(\begin{array}{cc}
0 & 1 \\
1 & 0
\end{array}\right) ,\qquad
\sigma_2 = \left(\begin{array}{cc}
0 & -i \\
i & 0
\end{array}\right) ,\qquad
\sigma_3 = \left(\begin{array}{cc}
1 & 0 \\
0 & -1
\end{array}\right).
$$  
Letting ${\mc S}_r$ be the totality of states $\rho_\theta$ of the above form, 
we have the hierarchy 
${\mc S}_1 \subset {\mc S}_2 \subset {\mc S}_3\subset \cdots \subset {\mc S}_n ={\mc S}$. 
Note that ${\mc S}_1$ is the set of product states 
$\rho_1 \otimes\rho_2 \otimes\cdots\otimes\rho_n$. 

Corresponding to the classical case, an element of 
${\mc S}_2$ is called a QBM (see \cite{NNjpa}).
The third-order quantum Boltzmann machines are given by the elements of 
 ${\mc S}_3$ and those states can be explicitly written as
\be
\rho_{h,w,v}=\exp\Bigl\{\sum_{i,s}
       h_{is} \sigma_{is} + 
\sum_{i < j}\sum_{s,t} w_{ijst} \sigma_{is} \sigma_{jt} + 
\sum_{i < j<k}\sum_{s,t,u} v_{ijkstu} \sigma_{is} \sigma_{jt} \sigma_{jt}
-\psi(h,w,v)\Bigr\}
\label{eq:3rdorderqbm} 
\ee
with 
\be 
\psi(h,w,v) = \log \Tr\exp\Bigl\{\sum_{i,s}
       h_{is} \sigma_{is} + 
\sum_{i < j}\sum_{s,t} w_{ijst} \sigma_{is} \sigma_{jt} + 
\sum_{i < j<k}\sum_{s,t,u} v_{ijkstu} \sigma_{is} \sigma_{jt} \sigma_{jt}\Bigr\}, 
\ee
where $h=(h_{is}),w=(w_{ijst})$ and $v=(v_{ijkstu})$.
\section{Some information geometrical concepts for quantum systems}
We discuss in this section some information geometrical concepts for quantum systems \cite{ANamsox}. Let us consider a manifold $\mc S$ of density operators and a submanifold $\mc M$ of $\mc S$. We define a quantum divergence function from $\rho\in\mc S$ to $\sigma \in \mc S$, which in this case turns out to be the quantum relative entropy and its reverse represented by
\be 
D^{(-1)}(\rho\|\sigma) \defeq \Tr[\rho(\log\rho - \log\sigma)];\;\;
D^{(+1)}(\rho\|\sigma) \defeq \Tr[\sigma(\log\sigma - \log\rho)].
\ee
The quantum relative entropy satisfies 
$D^{(\pm 1)}(\rho\|\sigma) \geq 0,\qquad D(\rho\|\sigma)=0\quad \mbox{iff}\quad 
\rho=\sigma$ but it is not symmetric. 

Given $\rho\in\mc S$, the point $\tau^{(\pm 1)} \in\mc M$ is called the e, m-projection of $\rho$ to $\mc M$, when function $D^{(\pm 1)}(\rho\|\tau),\; \tau\in\mc M$ takes a critical value at $\tau^{(\pm 1)}$, that is
\be
\frac{\pd}{\pd \xi} D^{(\pm 1)}(\rho\|\tau(\xi)) =0
\ee
at  $\tau^{(\pm 1)}$ where $\xi$ is a coordinate system of $\mc M$. the minimizer of $D^{(\pm 1)}(\rho\|\tau),\; \tau\in\mc M$, is the $\pm 1$-projection of $\rho$ to $\mc M$.
 
Next we introduce a quantum version of exponential family 
(\ref{eq:cexpfam}) in the following. 
Suppose that a parametric family 
\be
{\mc M} = \{\rho_{\theta}\:\vert\:\theta = (\theta^i);\,\, i=1,\ldots,m \}\,
\subset {\mc S}
\ee
 is represented in the form 
\be
\rho_{\theta} = \exp\Bigl\{C+\sum_{i=1}^m {\theta}^i F_i 
-\psi(\theta)\Bigr\},\label{eq:qexp}
\ee
where $F_i \,(i=1,\ldots,m),C$ are Hermitian operators and $\psi(\theta)$ is 
a real-valued function. We assume in addition that the operators 
$\{F_1,\ldots,F_m,I\}$, where $I$ is the identity operator, are linearly 
independent to ensure that the parametrization $\theta\mapsto\rho_\theta$ is 
one to one. Then ${\mc M}$ forms an $m$-dimensional smooth manifold with a 
coordinate system $\theta=(\theta^i)$. In this thesis, we call such an 
${\mc M}$ a quantum exponential family or QEF for short, with natural coordinates $\theta=(\theta^i)$. Note also that for any $1\leq k\leq n$ the set 
${\mc S}_k$ of states (\ref{eq:kelement}) forms a QEF, including ${\mc S}_1$ of product states, ${\mc S}_2$ of QBMs and ${\mc S}_3$ of third-order QBMs. 

If we let 
\be
\eta_i(\theta)\defeq \Tr\bigl[\rho_{\theta}F_i\bigr],
\label{eq:eta_general}
\ee
then $\eta = (\eta_i)$ and $\theta =(\theta^i)$ are in one-to-one 
correspondence. That is, we can also use $\eta$ instead of $\theta$ to specify 
an element of ${\mc M}$. These $(\eta_i)$ are called the expectation 
coordinates of ${\mc M}$. 

In particular, the natural coordinates of ${\mc S}_3$ are given by 
$(h,w,v) = (h_{is}, w_{ijst},v_{ijkstu})$ in (\ref{eq:3rdorderqbm}), while the expectation coordinates are $(m, \mu,\iota) = (m_{is}, \mu_{ijst},\iota_{ijkstu})$ defined by
\be
m_{is} = \Tr[\rho_{h,w,v}\,\sigma_{is}]\quad 
\mbox{and}\quad \mu_{ijst} = \Tr[\rho_{h,w,v}\,\sigma_{is} \sigma_{jt}] \mbox{and}\quad \iota_{ijkstu} = \Tr[\rho_{h,w,v}\,\sigma_{is} \sigma_{jt} \sigma_{ku}].
\label{eq:eta_3QBM}
\ee
On the other hand, the natural coordinates of ${\mc S}_1$ are 
$\bar{h}= (\bar{h}_{is})$ in (\ref{eq:prodstates}), while the expectation 
coordinates are $\bar{m}= (\bar{m}_{is})$ defined by 
\be
{\bar{m}}_{is} = \Tr[\tau_{\bar h} \sigma_{is}]. 
\label{eq:eta_product}
\ee
In this case, the correspondence between the two coordinate systems can 
explicitly be represented as 
\be
{\bar{m}}_{is} = \frac{\pd\psi_i(\bar{h}_i)}
{\pd {\bar{h}}_{is}} 
=\frac{{\bar{h}}_{is}}{||{\bar{h}}_i||}\tanh(||{\bar{h}}_i||) 
\label{eq:nmf1} 
\ee
 or as
\be
{\bar{h}}_{is} 
= \frac{{\bar{m}}_{is}}{||{\bar{m}}_i||}\tanh^{-1}(||{\bar{m}}_i||), 
\label{eq:nmf1inv} 
\ee
where $||{\bar{m}}_i|| \defeq \sqrt{\sum_s ({\bar{m}}_{is})^2}$. 

\section{Information geometry of mean-field approximation for third-order QBMs}
\subsection{The submanifold of product states and its geometry}
In this section, we briefly discuss the set ${\mc S}_1$. 
The elements of ${\mc S}_1$ are represented as $\tau_{h}$ by letting $w=0$ and $v=0$ in (\ref{eq:3rdorderqbm}). In the sequel, we write them as
\be
\tau_{\bar h} = \exp\Big\{\sum_{i,s} {\bar{h}}_{is} 
\sigma_{is} - \psi(\bar h)\Big\} \label{eq:prodstates}
\ee
by using new symbols $\tau$ and $\bar{h}= ({\bar{h}}_{is})$ when we wish to 
make it clear that we are treating ${\mc S}_1$ instead of ${\mc S}_3$. We have
\be
\tau_{\bar h} = \bigotimes_{i=1}^n \exp\Big\{\sum_s {\bar{h}}_{is} 
\sigma_s - \psi_i(\bar{h}_i)\Big\}, 
\label{eq:prodstates_var}
\ee
where $\bar{h}_i = ({\bar{h}}_{is})_s$ and
\bea
\psi_i(\bar{h}_i)  &=& 
\log\Tr\exp\Big\{\sum_{s} h_{is} \sigma_s \Big\}\nonumber\\ 
&=& \log\{ \exp(||{\bar{h}}_i||) + 
\exp(- ||{\bar{h}}_i||)\}\label{eq:psi1}
\eea
with $||{\bar{h}}_i|| \defeq \sqrt{\sum_s ({\bar{h}}_{is})^2}$. Note that 
\be 
\psi(\bar h) = \sum_i \psi_i(\bar{h}_i).\label{eq:psi2}
\ee
\subsection{The exponential \& mixture projections and mean-field approximation}
In this section, we derive the naive  mean-field equation for third-order QBMs 
explicitly from the viewpoint of information geometry. 
Suppose that we are interested in calculating the expectations 
$m_{is}=\Tr [\rho_{h,w} X_{is}]$ from given $(h, w) =(h_{is}, w_{ijst})$. 
Since the direct calculation is intractable in general  when the system size 
is large, we need to employ a computationally efficient approximation method. 
Mean-field approximation is a well-known technique for this purpose. 
The simple idea behind the mean-field approximation for a 
$\rho_{h,w,v}\in {\mc S}_3$ is to use quantities obtained in the form of 
expectation with respect to some relevant $\tau_{\bar h} \in {\mc S}_1$. 
T. Tanaka \cite{Tnc} has elucidated the essence of the naive 
mean-field approximation for classical spin models in terms of 
e-, m-projections. Our aim is to extend this idea to quantized spin models other than that considered in \cite{NNjpa}. 

In the following arguments, we regard ${\mc S}_3$ as a QEF with the natural 
coordinates $(\theta^\alpha) = (h_{is}, w_{ijst}, v_{ijkstu})$ and the expectation 
coordinates $(\eta_\alpha) = (m_{is}, \mu_{ijst}, \iota_{ijkstu})$, 
where $\alpha$ is an index denoting $\alpha=(i,s)$ or $\alpha=(i,j,s,t)$ or $\alpha=(i,j,k,s,t,u)$. 
We follow a slightly different method to that of classical setting to obtain naive mean-field equation for third-order QBM.

Recall that the state for third-order QBM (\ref{eq:3rdorderqbm}) is given by
\be
\rho_{h,w,v}=\exp\Bigl\{\sum_{i,s}
       h_{is} \sigma_{is} + 
\sum_{i < j}\sum_{s,t} w_{ijst} \sigma_{is} \sigma_{jt} + 
\sum_{i < j<k}\sum_{s,t,u} v_{ijkstu} \sigma_{is} \sigma_{jt} \sigma_{jt}
-\psi(h,w,v)\Bigr\} 
\ee
with 
\be 
\psi(h,w,v) = \log \Tr\exp\Bigl\{\sum_{i,s}
       h_{is} \sigma_{is} + 
\sum_{i < j}\sum_{s,t} w_{ijst} \sigma_{is} \sigma_{jt} + 
\sum_{i < j<k}\sum_{s,t,u} v_{ijkstu} \sigma_{is} \sigma_{jt} \sigma_{kt}\Bigr\}, 
\ee
where $h=(h_{is}),w=(w_{ijst})$ and $v=(v_{ijkstu})$.

Given $\rho_{h,w,v}\in {\mc S}_3$, its e-(+1) and m-(-1) 
projections (see \cite{ANamsox}) onto ${\mc S}_1$ are defined by 
\be
{\bar{\tau}}^{(\pm 1)} = \tau_{\bar{h}^{(\pm 1)}}\defeq 
\argmin_{\tau_{\bar{h}} \in {\mc S}_1} D(\rho_{h,w,v}\|\tau_{\bar{h}}).\label{eq:emproj1QBM}
\ee
We denote by ${\bar{m}}_{is}^{(\pm 1)}[\rho_{h,w,v}]$ the expectation of $\sigma_{is}$ with respect to ${\bar{\tau}}^{(\pm 1)}$, that is $\Tr[{\bar{\tau}}^{(\pm 1)} \sigma_{is}]$. Then ${\bar{m}}_{is}^{(\pm 1)}[\rho_{h,w,v}]$ is given by
\be
\frac{\pd}{\pd {\bar{m}}_is} D^{(\pm 1)}(\rho_{h,w,v}\|\tau_{\bar{h}}) =0. 
\ee
From the information geometrical point of view, ${\bar{\tau}}^{(\pm 1)} \in {\mc S}_1$ is the $\pm 1$-geodesic projection of $\rho_{h,w,v}$ to ${\mc S}_1$ in the sense that the $\pm 1$-geodesic connecting $\rho_{h,w,v}$ and $\bar{\tau}$ is orthogonal to ${\mc S}_1$ at $\bar{\tau}={\bar{\tau}}^{(\pm 1)}$. but we know that ${\bar{\tau}}^{(-1)}$ is $m$-projection of $\rho_{h,w,v}$ to ${\mc S}_1$. then we have ${\bar{m}}_{is}^{(-1)}{\bar{m}}_{is}[\rho_{h,w,v}]$ which is the quantity we want to obtain. This relation can be directly calculated by solving 
\be
\frac{\pd}{\pd {\bar{m}}_is} D^{(-1)}(\rho_{h,w,v}\|\tau_{\bar{h}}) =0, 
\ee
because this is equivalent to 
\be
\frac{\pd}{\pd {\bar{m}}_is} \Tr[\rho_{h,w,v}(\log \rho_{h,w,v}-\log\bar{\tau})] \frac{\pd}{\pd {\bar{m}}_is} \Tr[ \rho_{h,w,v}\log\bar{\tau})]=0. 
\ee
Hence ${\bar{m}}_{is}^{(-1)} = \Tr[\rho_{h,w,v} \sigma_{is}]$ which is the quantity we have been searching for. But we cannot calculate $\Tr[\rho_{h,w,v} \sigma_{is}]$ explicitly due to the difficulty in calculating $\psi(h,w,v)$ for $\rho_{h,w,v}$.

If we use the $e$-projection of $\rho \in {\mc S}_3$ to ${\mc S}_1$ instead of the $m$-projection, we have the naive mean-field approximation as in the classical case. to show this we calculate the $e$-projection (1-projection) of $\rho$ to ${\mc S}_1$. Then we have
\bea
D^{(1)} (\rho\|\tau) &=& D^{(-1)} (\tau\|\rho) = \Tr[\tau(\log\tau-\log\rho)]\\
&=& \Tr\biggl[\tau\biggl\{\biggl(\sum_{i,s} {\bar{h}}_{is} 
\sigma_{is} - \psi(\bar h)  \biggr) \nn\\
&&- \biggl( \sum_{i,s}
       h_{is} \sigma_{is} + 
\sum_{i < j}\sum_{s,t} w_{ijst} \sigma_{is} \sigma_{jt} + 
\sum_{i < j<k}\sum_{s,t,u} v_{ijkstu} \sigma_{is} \sigma_{jt} \sigma_{jt}
-\psi(h,w,v) \biggr)  \biggr\}  \biggr]\\
&=& \sum_{i,s} {\bar{h}}_{is} {\bar{m}}_{is} - \psi(\bar h) - \sum_{i,s} h_{is} {\bar{m}}_{is}
+ \sum_{i < j}\sum_{s,t} w_{ijst} {\bar{m}}_{is} {\bar{m}}_{jt} \nn\\
&&+ 
\sum_{i < j<k}\sum_{s,t,u} v_{ijkstu} {\bar{m}}_{is} {\bar{m}}_{jt} {\bar{m}}_{ku}-\psi(h,w,v),
\eea
where we define ${\bar{m}}_{is} = \Tr[\tau \sigma_{is}], 
{\bar{m}}_{is} {\bar{m}}_{jt}= \Tr[\tau \sigma_{is}\sigma_{jt}]$ and 
${\bar{m}}_{is} {\bar{m}}_{jt} {\bar{m}}_{ku}= \Tr[\tau \sigma_{is}\sigma_{jt}\sigma_{ku}]$. Hence

\bea
\frac{\pd}{\pd {\bar{m}}_is} D^{(1)} &=& 
\frac{\pd}{\pd {\bar{m}}_is} \biggl[\sum_{i,s} {\bar{h}}_{is} {\bar{m}}_{is} - \psi(\bar h)  \biggr] -  h_{is} 
+ \sum_{j\neq i}\sum_{s,t} w_{ijst} {\bar{m}}_{jt} + 
\sum_{k \neq j \neq k}\sum_{s,t,u} v_{ijkstu} {\bar{m}}_{jt} {\bar{m}}_{ku}\\
&=& {\bar{h}}_{is} -  h_{is} 
+ \sum_{j\neq i}\sum_{s,t} w_{ijst} {\bar{m}}_{jt} + 
\sum_{k \neq j \neq k}\sum_{s,t,u} v_{ijkstu} {\bar{m}}_{jt}{\bar{m}}_{ku} = 0.
\eea
This gives
\be
{\bar{h}}_{is} =  h_{is} + \sum_{j\neq i}\sum_{s,t} w_{ijst} {\bar{m}}_{jt} + 
\sum_{k \neq j \neq k}\sum_{s,t,u} v_{ijkstu} {\bar{m}}_{jt}{\bar{m}}_{ku}
\label{eq:nmf3qbm1}
\ee
and ${\bar{m}}_{is}$ is given by
\be
{\bar{m}}_{is} = \frac{\pd\psi_i(\bar{h}_i)}
{\pd {\bar{h}}_{is}} 
=\frac{{\bar{h}}_{is}}{||{\bar{h}}_i||}\tanh(||{\bar{h}}_i||).
\label{eq:nmf3qbm2} 
\ee
Both (\ref{eq:nmf3qbm1}) and (\ref{eq:nmf3qbm2}) together give the naive mean-field equations for third-order QBMs.
\section{Concluding remarks}
We have applied information geometry to the mean-field approximation for a general class of quantum statistical models. Here, we were able to derive only the naive mean-field equations. However, it is known that the naive mean-field approximation does not give a good approximation to the true value. Therefore, to improve the approximation we need to consider the higher order approximations and the information geometrical point of view is left open.


\begin{thebibliography}{9}
\bibitem{NNjpa} N.~Yapage, H.~Nagaoka.
An information geometrical approach to the mean-field approximation for 
quantum Ising spin models. {\it J. Phys. A: Math. Theor.} {\bf 41} (2008) 
065005.

\bibitem{ANamsox} S. Amari and H. Nagaoka.
{\it Methods of Information Geometry}. American Mathematical Society and 
Oxford University Press, 2000.

\bibitem{MDmit} Manfred Opper and David Saad. 
{\it Advanced Mean Field Methods - Theory and Practice}, 
MIT Press, Cambridge, MA, 2001.

\bibitem{Tnc} T.~Tanaka,
Information Geometry of mean field Approximation. 
{\it Neural Computation}, {\bf 12} pp.1951--1968, 2000.

\bibitem{AKNnn} S. Amari, K. Kurata and H. Nagaoka. 
Information Geometry of Boltzmann Machines. {\it IEEE Trans. on Neural
Networks} Vol 3, No. 2, pp.260--271, 1992.

\bibitem{TJS} T. ~J. Sejnowski. Higher-order Boltzmann Machines. 
{\it Conference Proceedings 151: Neural Networks for Computing}(Snowbird, Utah) 1986.

\bibitem{NKcs} H. Nagaoka and T. Kojima. Boltzmann Machine as a 
Statistical Model. 
{\it Bulletin of The Computational Statistics of Japan} Vol.1, 
pp.61--81, 1995 (in Japanese).


\end{thebibliography}
\end{document}